\documentclass[pmlr]{jmlr}
\usepackage[normalem]{ulem}
\usepackage{wrapfig}

\RequirePackage{graphicx}
 \usepackage{booktabs}
\usepackage{longtable}
\makeatletter
\def\set@curr@file#1{\def\@curr@file{#1}} 
\makeatother
\usepackage[load-configurations=version-1]{siunitx} 

\makeatletter

\usepackage{comment}
\let\wfs@comment@comment\comment
\let\comment\@undefined
\usepackage{changes}
\let\wfs@changes@comment\comment
\let\comment\@undefined
\newcommand\comment{%
    \ifthenelse{\equal{\@currenvir}{comment}}
    {\wfs@comment@comment}
    {\wfs@changes@comment}%
}
\makeatother

\usepackage{amsmath}
\usepackage{hyperref}
\usepackage{enumitem}
\usepackage{multirow}
\usepackage{outlines}
\usepackage{graphicx}
\usepackage{algorithm}
\usepackage{float}
\usepackage{nicefrac}
\usepackage{url}
\usepackage{booktabs}
\usepackage{listings}
\usepackage{tcolorbox}
\tcbuselibrary{listings, breakable}

\usepackage{wrapfig}
\usepackage{pifont}

\definechangesauthor[name="Ilya Tyagin", color=teal]{it}
\definechangesauthor[name="Ilya Safro", color=red]{is}
\definechangesauthor[name="Saeideh Valipour", color=green]{sv}

\theorembodyfont{\upshape}
\theoremheaderfont{\scshape}
\theorempostheader{:}
\theoremsep{\newline}

\usepackage{enumitem}

\jmlrvolume{298}
\jmlryear{2025}
\jmlrworkshop{Machine Learning for Healthcare}

\title[Biomedical Hypothesis Explainability with Graph-Based Context Retrieval]{Biomedical Hypothesis Explainability with Graph-Based Context Retrieval}

\author{\Name{Ilya Tyagin}
       \Email{tyagin@udel.edu}\\ 
       \addr Center for Bioinformatics and Computational Biology\\
       University of Delaware\\
       Newark, DE, USA 
       \AND
       \Name{Saeideh Valipour}
       \Email{svalipou@udel.edu}\\ 
       \addr Computer and Information Sciences\\
       University of Delaware\\
       Newark, DE, USA
       \AND
       \Name{Aliaksandra Sikirzhytskaya}
       \Email{sikirzha@mailbox.sc.edu}\\ 
       \addr  Drug Discovery and Biomedical Sciences (DDBS) College of Pharmacy\\
       University of South Carolina\\
       Columbia, South Carolina, USA 
       \AND
       \Name{Michael Shtutman}
       \Email{shtutmanm@cop.sc.edu}\\ 
       \addr  Drug Discovery and Biomedical Sciences (DDBS) College of Pharmacy\\
       University of South Carolina\\
       Columbia, South Carolina, USA 
       \AND
       \Name{Ilya Safro}
       \Email{isafro@udel.edu}\\ 
       \addr  Computer and Information Sciences\\
       University of Delaware\\
       Newark, DE, USA
       } 

\begin{document}

\maketitle
\vspace{-5em}
\begin{abstract}
We introduce an explainability method for biomedical hypothesis generation systems, built on top of the novel Hypothesis Generation Context Retriever framework. Our approach combines semantic graph-based retrieval and relevant data-restrictive training to simulate real-world discovery constraints. Integrated with large language models (LLMs) via retrieval-augmented generation, the system explains hypotheses with contextual evidence using published scientific literature. We also propose a novel feedback loop approach, which iteratively identifies and corrects flawed parts of LLM-generated explanations, refining both the evidence paths and supporting context. We demonstrate the performance of our method with multiple large language models and evaluate the explanation and context retrieval quality through both expert-curated assessment and large-scale automated analysis. Our code is available at:~\url{https://github.com/IlyaTyagin/HGCR}.
\end{abstract}

\section{Introduction}
Automated biomedical hypothesis generation (HG, also known as Literature-Based Discovery), aims to uncover implicit biomedical connections from large-scale literature corpora. Originating from Swanson’s idea of “undiscovered public knowledge” \citep{swanson1986undiscovered}, HG identifies non-trivial links between biomedical concepts to support testable scientific insights \citep{popper59logic}. Over time, HG systems have evolved from simple keyword overlap methods to those leveraging knowledge graphs and deep learning \citep{henry2017literature, cesario2024survey}.

However, most HG systems often lack interpretability, offering unexplainable numeric predictions which limits their practical utility and weakens user trust. Integrating HG with predictive modeling can improve biomedical discoveries (e.g., identifying gene-disease links \citep{sybrandt2018large,aksenova2020inhibitionfull, cummings2022exposure}), but explainability remains a critical bottleneck.

Recent work explores Large Language Models (LLMs) for HG explainability. However, LLMs, while performing well for language generation, often hallucinate facts and lack domain-specific reasoning, leading to scientifically implausible hypotheses. They are also unaware of temporal constraints and often fail to align with structured biomedical knowledge.

To address this, we propose the Hypothesis Generation Context Retrieval (HGCR), a retrieval-augmented framework for hypothesis generation with explainability. 
HGCR constructs a dynamic biomedical co-occurrence graph from MedLine abstracts, where nodes represent biomedical concepts using the Universal Medical Language System (UMLS CUIs, \citep{bodenreider2004unified}) and edges represent co-occurrences of these concepts in abstracts. The system then identifies and ranks semantic paths between concept pairs in this graph based on their plausibility. Retrieved path is associated with the specific Medline abstracts/literature where the concepts co-occur, creating path-literature pairs and used as context for LLM-generated explanations.

To ensure scientific soundness of generated hypothesis explanations, we introduce a feedback loop architecture that uses AGATHA \citep{agathasybrandt20} and SemRep \citep{rindflesch03} to extract and validate semantic predicates (i.e., subject–verb–object triples) representing biomedical relationships between UMLS terms from the LLM output. Unsupported claims trigger context update and iterative explanation refinement.

\paragraph{Our contribution:}
\begin{enumerate}[label=(\arabic*), leftmargin=*, itemsep=0em]
    \item We propose HGCR, a graph-based biomedical context retrieval system that constructs semantic paths from the whole MEDLINE dataset of biomedical abstracts and citations to explain potential relationships between concepts of interest.
    \item We introduce a feedback mechanism that validates LLM-generated explanations using biomedical hypothesis generation system (in our experiments we use AGATHA) by iteratively refining context to minimize the number of LLM-generated semantic predicates identified by HG system as potentially wrong.
    \item We perform a retrospective temporal evaluation of both retrieval and explanation quality, benchmarking HGCR and the feedback loop pipeline against existing retrieval-augmented generation (RAG) systems and demonstrate the performance of the proposed method by human expert evaluation.
\end{enumerate}

\subsection*{Generalizable Insights about Machine Learning in the Context of Healthcare}
Our study presents key insights for developing more effective ML systems in healthcare workflows. First, automatic explainability is crucial especially in domains like biomedicine, where interpretable reasoning behind predictions is as important as their accuracy and clinical adoption. Second, the combination of knowledge graphs into retrieval-augmented generation frameworks shows promise, as integrating LLMs with structured knowledge and external validation tools (like AGATHA) results in more scientifically sound hypotheses than using LLMs alone. Finally, iterative refinement based on feedback and validated evidence enhances the plausibility and utility of ML-generated hypotheses, highlighting socio-technical benefits relevant for deployment in real-world biomedical settings, usefulness of generated hypotheses and their explanations.
 

\section{Related Work}
\paragraph{Hypothesis Generation.}
Scientific HG was pioneered by Swanson  \citep{swanson1986undiscovered} who proposed to discover unknown connections by reasoning over semantically disconnected literature. Subsequent systems evolved to graph-based and topic modeling methods (e.g., MOLIERE \citep{sybrandt2017moliere} or BioLDA \citep{wang2011finding}, which applied LDA topic modeling and used multi-modal semantic graphs to extract plausible biomedical associations.
Many frequency-based approaches such as co-occurrence frequency of term \citep{srinivasan2004text} other frequencies such as relative frequency \citep{lindsay1999literature}, tf-idf \citep{srinivasan2004text} were also investigated to complement Swanson work. The high co-occurrence frequencies do not necessarily guarantee meaningful relationships. More recently, a series of text mining technologies, such as random walking \citep{shi2015weaving}, Latent Semantic Indexing (LSI) \citep{gordon1998using}, ranking \citep{rastegar2015new}, association rules \citep{yetisgen2006using} have been investigated to learn more implicit semantics of terms, inferring complex semantic associations. Despite the improvements in modeling semantic associations, graph theoretic machine learning approaches have also made swift inroads into HG task for complex association generation, such as constructed a subgraph to provide for deep understanding of the associations \citep{cameron2015context}, extracting the graph pattern features from graphs to infer treatment and causative relations \citep{bakal2018exploiting}, applying the direct edge searches and meta-paths in knowledge graphs (KG) for hypothesis generation \citep{taneja2023developing}. While biomedical literature is growing exponentially with new knowledge being discovered every day, The temporal dynamics of scientific term relations has been studied by several recent works \citep{jha2019hypothesis,xun2017generating,akujuobi2020tpair,zhou2022learning}.

\paragraph{AGATHA.}
AGATHA \citep{agathasybrandt20,tyagin22} introduced a transformer-based hypothesis generation pipeline over a structured semantic graph derived from MEDLINE, producing link plausibility scores. Later works explored generative approaches, such as BioGPT \citep{luo22} and CBAG \citep{sybrandt2021cbag}, as well as recent attempts to directly use LLMs for hypothesis formulation \citep{qi2024largelanguagemodelsbiomedical,iser2024automatedexplanationselectionscientific}. While promising, LLM-only methods often hallucinate connections lacking mechanistic or temporal validity, that is, they do not correspond to known biological or causal processes. This motivates retrieval-augmented or hybrid approaches, such as RUGGED \citep{pelletier2024explainablebiomedicalhypothesisgeneration} and the proposed HGCR framework, which use graph-structured paths to improve hypotheses based on plausible biomedical context.

\paragraph{LLMs and Transformers in Biomedical Research.}
Biomedical-specific LLMs such as BioBERT \citep{lee2019biobert}, PubMedBERT \citep{gu2021pubmedbert}, and SciFive \citep{phan21} enable named entity recognition, question answering, and classification on MEDLINE scale corpora. However, generative models like PaLM \citep{chowdhery22} or GPT-4, while capable of synthesizing plausible hypotheses, frequently lack interpretability and scientific validity \citep{elbadawi24,park24}. HGCR addresses this limitation by treating hypothesis explanation as a path ranking problem, enriched with literature-derived context and embeddings. Furthermore, it introduces a feedback loop for explanation refinement, correcting unsupported claims through external validation with hypothesis generation systems like AGATHA.

\paragraph{Biomedical Question Answering (QA) and RAG.}
Medical QA systems have progressed from BERT-based models \citep{devlin19,jin22} to LLMs fine-tuned on biomedical datasets \citep{singhal23,wu23}. RAG architectures \citep{zuheros2021sentiment} mitigate hallucination by providing relevant documents to generate LLM responses. In the biomedical domain, systems like Clinfo.ai \citep{lozano23}, BiomedRAG \citep{li2024biomedragretrievalaugmentedlarge}, and MedRAG \citep{xiong24} retrieve MEDLINE abstracts to support QA or relation extraction. iMedRAG \citep{xiong24biMedRag} adds iterative query refinement for multi-step reasoning.

The proposed HGCR framework complements these approaches by ranking semantically coherent graph paths rather than flat documents, enabling structured retrieval of mechanistic explanations. Unlike existing RAG systems, it incorporates a contrastive training objective to distinguish meaningful reasoning chains from corrupted or unrelated paths, and includes a refinement loop to iteratively align generated explanations with contextual evidence.

\section{Methods}
The schema of the proposed hypothesis self-adjusting algorithm (called {\textit{feedback loop}) is shown in Figure~\ref{fig:flow}. It integrates the biomedical knowledge network, context retriever, LLMs and automatic evaluator to generate and refine indirect explanations between biomedical concept pairs. It bridges the gap between machine-generated predictions and human interpretability by converting ranked entity relationships into natural language descriptions supported by literature evidence.

\begin{figure}[t]
 \centering 
 \includegraphics[width=6.5in]{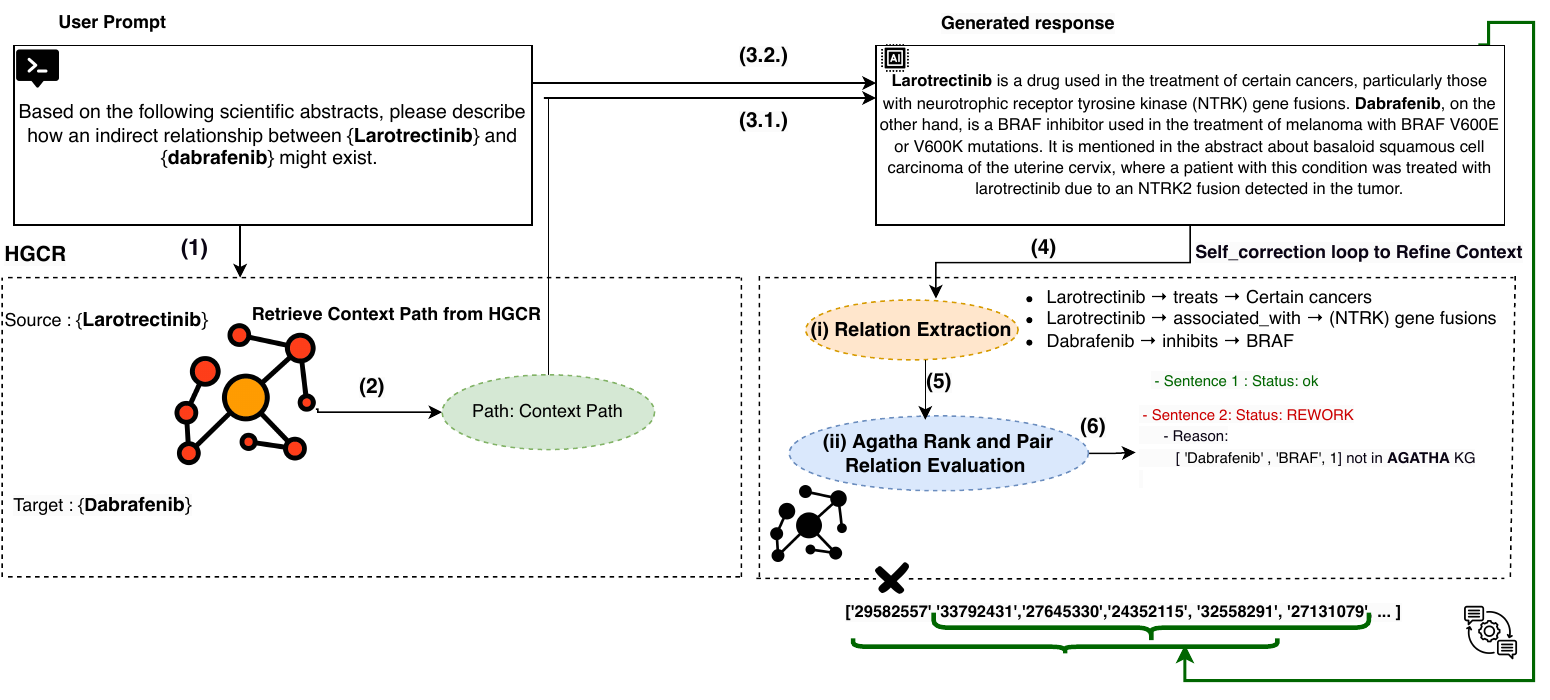} 
 \caption{
 Overview diagram of the proposed pipeline based on Hypothesis Generation Context Retriever (HGCR) and self-correction loop for context refinement.
 } 
 \label{fig:flow} 
\end{figure} 
\subsection{Context Path Retrieval with HGCR}
\paragraph{Graph structure and temporal context.}

The system operates over a dynamic biomedical semantic network  \( G = \{G_t\}_{t=1}^T \), where each snapshot represents the state of the network at time \( t \):
\begin{equation}
G_t = (N_t, E_t).
\label{eq:graph_snapshot}
\end{equation}

Nodes $N_t$ correspond to unique UMLS Concept Unique Identifiers (CUIs), and edges $E_t$ represent validated biomedical associations between them~\citep{dyport}. An association is defined as the co-occurrence of two UMLS concepts within the same MEDLINE abstract, without requiring them to appear within the same sentence. Each edge $(u, v) \in E_t$ is supplied with a set of PubMed identifiers (PMIDs) referencing biomedical literature that supports the association and a corresponding publication year $t$.

Temporal evolution is defined by cumulative literature support: for each year $t$, the edge set $E_t$ includes all associations observed in the literature up to that year  with PMIDs that correspond to the paper abstract:

\begin{equation}
    E_t = \{(u, v) \mid \exists\, \text{PMID associated with } (u, v) \text{ and } \text{year} \leq t\}
\end{equation}

The graph dynamically grows as new associations are discovered and published, with subsequent snapshots $G_t$ expanding upon earlier ones. This structure allows for modeling temporal biomedical knowledge with fine-grained control over historical context.

\paragraph{Path-finding and dataset generation.}

A path $p_k$ is defined within a specific snapshot \( G_t = (N_t, E_t) \) of the dynamic graph \( G = \{G_t\}_{t=1}^T \). It consists of an ordered sequence of nodes connecting a source node \( m_i \in N_t \) to a target node \( m_j \in N_t \), traversing intermediate nodes \( m_{k_1}, m_{k_2}, \ldots, m_{k_n} \in N_t \), such that each consecutive pair of nodes in the sequence is connected by an edge in \( E_t \). Formally, a path is defined as:
\begin{equation}
p_k = (m_i, m_{k_1}, m_{k_2}, \ldots, m_{k_n}, m_j), \quad \text{with } (m_{k_\ell}, m_{k_{\ell+1}}) \in E_t \text{ for all } \ell
\end{equation}
Additionally, each edge $(m_{k_i}, m_{k_j})$ within the path is associated with one or more PMIDs, indicating the source of its literature context, which is an integral part of the overall path representation.

\paragraph{Positive and negative samples.}

For each source-target pair $(m_i, m_j)$, where a validated association appears at timestamp $t$, we construct positive training samples by identifying sets of paths by finding shortest paths from the network $G$ at timestamp $t-1$. 

To generate positive samples, we extract a set of future reference terms $\mathcal{F}_{(m_i, m_j)}^t$ from PubMed abstracts published at timestamp $t$ that describe the association $(m_i, m_j)$. Specifically, we collect all terms $m_{q}$ mentioned in the abstracts associated with the PMIDs that refer to the discovery $(m_i, m_j)$ at timestamp $t$:

\begin{equation}
    \mathcal{F}_{(m_i, m_j)}^t= \{ m_q \mid \text{PMID}(m_i, m_j) \text{ at } t \text{ mentions } m_q\}
\end{equation}

We define a set of positive paths \( \mathcal{P}^+ \). A path \( p_k = (m_i, m_{k_1}, m_{k_2}, \ldots, m_{k_n}, m_j) \in \mathcal{P}^+ \) is labeled as positive if its set of intermediate nodes is contained in the set of future reference terms \( \mathcal{F}^t_{(m_i, m_j)} \). Formally:

\begin{equation}
p_k \in \mathcal{P}^+ \implies (m_{k_1}, m_{k_2}, \ldots, m_{k_n}) \subseteq \mathcal{F}^t_{(m_i, m_j)}
\end{equation}

The positive label indicates that the sequence of nodes in the path (and the associated context) includes the future reference terms, suggesting that this path is associated with the discovery. The main idea is that if intermediate concepts in a path appear in future scientific literature discussing the target association, then the path likely reflects a valid reasoning trajectory. 

We generate three types of negative samples to enhance the model’s discriminative ability.

\subparagraph{Hard Negative Samples.}
Hard negative samples are constructed by sampling paths from the network snapshot \( G_{t-1} \), which represents the state of the biomedical co-occurrence graph prior to the discovery at timestamp \( t \); they are intended to represent trajectories that are not related to the future discovery $(m_i, m_j)$. 

For a given source-target pair $(m_i, m_j)$, a path $p_k$ is labeled as negative if it includes intermediate node(s) that are not part of the future reference set $\mathcal{F}_{(m_i, m_j)}^t$:

\[
p_k \in P^{-}_{\text{hard}} \quad \Longrightarrow \quad \exists\, m_{k_r} \in \{m_{k_1}, m_{k_2}, \dots, m_{k_n}\} : m_{k_r} \notin \mathcal{F}^{t}_{(m_i, m_j)}
\]

Hard negatives \( P^-_{\text{hard}} \) are sampled from a pool of paths between \( m_i \) and \( m_j \) that are structurally valid (i.e., shortest paths in the co-occurrence graph) but were not part of future-relevant literature.

\subparagraph{Corrupted Paths.}
Corrupted paths \( P^-_{\text{corr}} \) are generated by \textit{introducing noise} into positive paths through \textit{node replacement}:

\[
P^-_{\text{corr}} = (m_i, m_{k_1}, \ldots, m_{k_{r-1}}, m_{k_r}', m_{k_{r+1}}, \ldots, m_j)
\]
where \( m_{k_r}' \) is a randomly sampled term from a valid set of terms $N_{t-1}$ such that \( m_{k_r}' \neq m_{k_r} \). This corruption simulates \textit{near-miss scenarios}, where only a single term is altered.

\subparagraph{Corrupted Contexts.}
Negative path samples with corrupted contexts \( P^-_{\text{corr\_ctx}} \) are constructed by pairing a structurally valid path \( p_k \in P^+ \) with deliberately simulated corrupted contextual information. Specifically, we retrieve PMIDs related to individual nodes of a positive path \( P^+ \), but \textit{without maintaining their co-occurrence relationships}:

\[
C^-_{\text{corr}} = \{ c_i \mid i \in P^+,\, c_i \in \mathcal{C}_{\text{node}} \}
\]
where \( \mathcal{C}_{\text{node}} \) refers to context abstracts retrieved by querying each node independently, breaking their relational structure.

We define the corrupted context path sample as:

\[
P^-_{\text{corr\_ctx}} = \left\{ (p_k, C^-_{\text{corr}}) \,\middle|\, p_k \in P^+,\, C^-_{\text{corr}} \not\subseteq C^+ \right\}
\]

This type of negative sampling forces the model to distinguish between truly meaningful sequences of evidence and those composed of isolated, unrelated abstracts, despite otherwise valid path structure.

As a result, for every positive sample $p_k \in P^+$ we have the following set of negative samples:
\[
P^- = P^-_{\text{hard}} \cup P^-_{\text{corr}} \cup P^-_{\text{corr\_ctx}}
\]

\subsection{Ranking Hypothesis Context}

\begin{figure}
    \centering
    \includegraphics[
        width=0.8\linewidth,
    ]{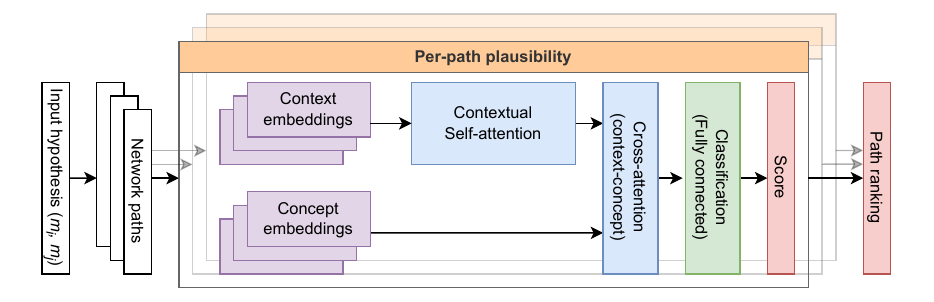}
    \caption{%
        \label{fig:hgcr_architecture}
        Overview of the proposed HGCR framework. 
        Input is a hypothesis: a pair of biomedical concepts $(m_i, m_j)$. The system samples paths from the network $G$ and ranks them based on their predicted alignment with the future reference context.
    }
\end{figure}

The proposed framework is designed to evaluate network paths between a given source term \( m_i \) and a target term \( m_j \). The goal is to score these paths based on their relevance to a future discovery, inferred from the past data. The model is trained using a contrastive learning approach, optimizing the ability to distinguish positive paths from various types of negative paths discussed earlier.

\subsubsection{Objective Function}

The model is trained using a margin ranking loss to encourage the model to assign higher scores to plausible (future-validating) context-path pairs than to negative (implausible or corrupted) ones. Given a score for a positive sample \( \hat{S}^+ \) and a set of scores for negative samples \( \{\hat{S}^-_i\}_{i=1}^N \), the loss is defined as:

\[
\mathcal{L} = \frac{1}{N} \sum_{i=1}^{N} \max\left(0, \delta - \left(\hat{S}^+ - \hat{S}^-_i\right)\right),
\]
where \( \delta \) is a predefined margin, which we set to $0.3$ for the best balance between classes separation and performance.

\subsubsection{Model Architecture}

The HGCR model computes a plausibility score for a candidate network path connecting two biomedical terms \( (m_i, m_j) \). 
It leverages two sources of information: the structural information encoded in concept embeddings \( P = \{\mathbf{e}_1, \dots, \mathbf{e}_k\} \), where each \( \mathbf{e}_i \in \mathbb{R}^{d_n} \) represents a sequence of \( k \) nodes in the path; and the contextual information encoded in abstract embeddings \( C = \{\mathbf{c}_1, \dots, \mathbf{c}_m\} \), where each \( \mathbf{c}_j \in \mathbb{R}^{d_p} \) is derived from a corresponding MEDLINE abstract. The model architecture is shown in Figure~\ref{fig:hgcr_architecture}.

The model first applies multi-head self-attention to the context embeddings to capture the interactions among different elements of the abstract context. The resulting context representations are then used as queries in a cross-attention block, while the concept embeddings are used as keys and values, such that each context position can attend to all path nodes. This produces a sequence-level path representation that integrates graph-based structural information into the contextual evidence. The resulting sequence is mean-pooled and passed through a fully connected classification layer with sigmoid activation to produce a final plausibility score \( \hat{S} \in [0, 1] \): 
\begin{align*}
C' &= \text{SelfAttn}(C) \\
A &= \text{CrossAttn}(Q = C',\ K = P,\ V = P) \\
\hat{S} &= \sigma(W \cdot \text{MeanPool}(A) + b)
\end{align*}

This architecture is conceptually related to cross-modal models such as LXMERT and VilBERT~\citep{tan2019lxmert, lu2019vilbert}, which use self-attention followed by cross-attention to align different input modalities. Similarly, HGCR aligns structured (node-based) and unstructured (text-based) information via stacked attention mechanism, and produces path-level predictions based on their joint representation.

At inference time, paths are ranked based on their scores and are then passed to a downstream pipeline along with their context to serve as a basis for hypothesis explainability framework. 

\subsection{Explainability Framework}
In this phase of the system, the goal is to transform latent, graph-based biomedical connections into interpretable explanations. To achieve this, we introduce a structured explainability pipeline that consists of four main components: 
(1) prompt construction with appended retrieved context from HGCR, 
(2) explanation generation via LLM, 
(3) validation with relation extraction and HG system ranking, and 
(4) iterative feedback loop for context refinement. 
The following subsections detail each component of this explainability pipeline, starting with how prompts are constructed and submitted to LLM for generating explanations.

\paragraph{Prompt Construction and LLM Generation.}
Step 1 in Figure~\ref{fig:flow} illustrates the prompt provided to LLM, which integrates scientific abstracts retrieved by HGCR as contextual input.
The prompt is framed as follows: 
``Based on the following scientific abstracts, please describe how an indirect relationship between \{SOURCE\} and \{TARGET\} might exist.'' 
In Step 2 we append context retrieved from HGCR to the prompt that is finally executed by LLM.

\paragraph{Structured Relation Extraction.}
Once LLM generates an explanation from the provided contextualized prompt (Figure~\ref{fig:flow}, Steps 3.1 and 3.2), the output is passed to a relation extraction module (Step 4), which parses the explanation into structured biomedical predicates using SemRep. 
Each predicate is extracted in the form of \textit{(subject, verb, object)}. 
All terms are normalized to UMLS Concept Unique Identifiers (CUIs) to ensure consistency and compatibility with downstream modules. 

The extracted predicates serve as the basis for validating the factual accuracy of the generated explanation. 
At this point, the system transitions from generation to evaluation and refinement (i.e., the feedback loop).

\paragraph{Relation Evaluation via AGATHA Evaluator.}  
\label{rel_eval_w_AGATHA}
Each extracted predicate is evaluated using the AGATHA Evaluator (Figure~\ref{fig:flow}, Step 5), which performs both direct validation and plausibility ranking. 

In the direct validation phase, the extracted predicate is matched against a precomputed set of biomedical relationships in the AGATHA semantic network. If no exact match is found, the system treats it as a hypothesis and computes its plausibility score using AGATHA predictor. This score is then compared against the scores of semantically related, but randomly generated biomedical concept pairs. The predicate is considered valid if its score ranks within the top 10\% among this set. Predicates failing both checks are flagged as scientifically implausible.
When this occurs, the corresponding sentence in the explanation is labeled as a \textit{rework sentence}. 
These sentences identify problematic parts in the explanation that lack scientific support and trigger the context refinement process described next.

\paragraph{Feedback Loop and Context Refinement.}

The feedback loop mechanism (Step 6) allows the system to iteratively refine the prompt context in response to rework sentences. 
For each flagged sentence, the system identifies which PMID in the current context most likely contributed to the unsupported relationship.

The system then uses MedCPT embeddings to embed all candidate abstracts (PMIDs) associated with the same edge as the flagged sentence. 
Cosine similarity between the rework sentence and each candidate abstract is computed, and the top-ranked abstract that has not been already used is selected as a replacement. 
This updated “rework-aware context” is then  used in the prompt, and LLM is queried again with the new context (returning to Step 3.1).

This feedback loop continues iteratively, with the explanation being regenerated, re-evaluated, and further refined as needed. 
The loop terminates either when all sentences pass validation or when a maximum number of iterations is reached (in our experiments $n = 5$). 
Each source-target path is refined independently, ensuring that explanations are tailored to their specific biomedical context.

\subsection{Baseline Approach}
The baseline experiment evaluates how well LLMs generate coherent, scientifically plausible explanations for indirect biomedical relationships using a fixed context without feedback loop. 
For each path \( p_k \), retrieved by the HGCR system, we collect abstracts for each edge and rank them by semantic similarity to the node pair of the corresponding edge using MedCPT embeddings.
Then an explanation is generated by prompting an LLM as follows:

``Based on the following scientific abstracts, please describe how an indirect relationship between \{SOURCE\} and \{TARGET\} might exist. Consider the key findings, underlying mechanisms, and any intermediate entities or processes mentioned in the abstracts. Your explanation should connect these elements to form a coherent narrative that illustrates the possible indirect linkage between \{source\} and \{target\}.''

\section{Experiments}

\paragraph{Experimental setup.} 
We describe the experimental setup for two key components: graph-based context retrieval (HGCR) and explainability framework.
The explainability framework is then run in two different settings: baseline and feedback loop.
The baseline experiment evaluates the performance of the entire system in generating hypotheses from fixed context retrieved by the HGCR system, while the feedback loop experiment iteratively refines the hypotheses by leveraging generated responses and incorporating feedback to improve the quality of the generated explanations.

\paragraph{Test dataset.}
\begin{figure}
    \centering
    \includegraphics[
        width=1\linewidth,
    ]{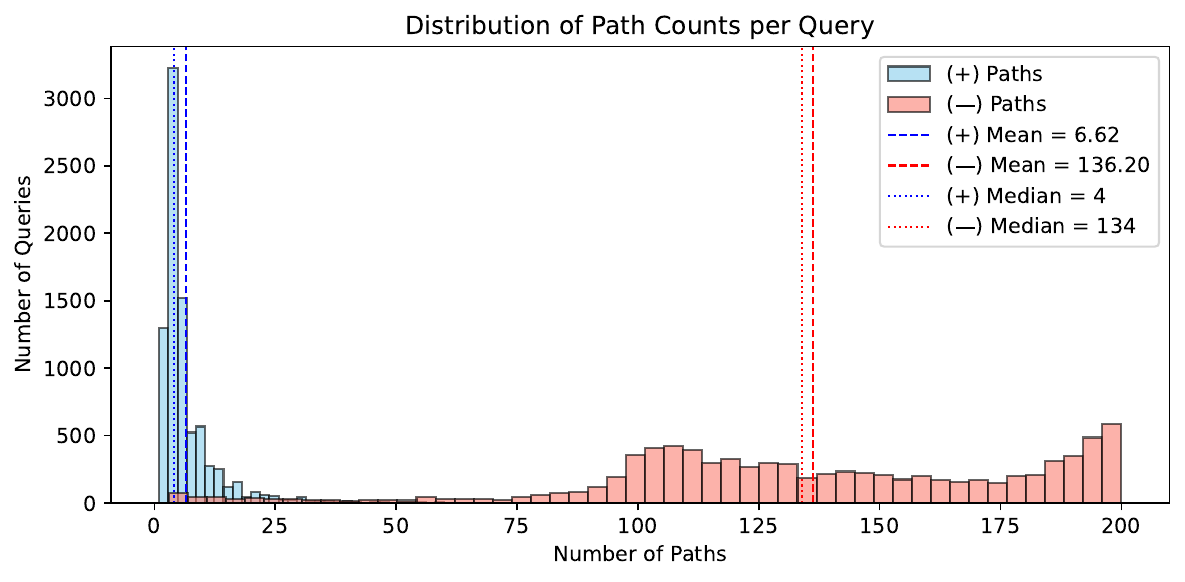}
    \caption{%
        \label{fig:test_paths_len_hist}
        Distribution of positive and negative path counts per query in the test set. Queries $q$ were extracted from the timestamp $t$ corresponding to 2022 (and onwards) and the shortest paths between them were sampled from $G_{t-1}$, representing the graph state of 2021. 
    }
\end{figure}

The evaluation of biomedical hypothesis generation and discovery-driven pipelines is challenging \citep{sybrandt2018large}, particularly in terms of assessing the quality of their explainability. Our evaluation methodology is an attempt to reflect on how scientific discoveries are made and validated over time.

Our test samples are extracted from the hypothesis generation benchmarking framework Dyport~\citep{dyport}. It provides a dynamic biomedical knowledge graph \( G_t \) (defined in Equation~\ref{eq:graph_snapshot}), where nodes $m_k \in N_t$ are UMLS concepts and edges \( (m_i, m_j) \in E_t \) represent validated biomedical discoveries at timestamp \( t \), annotated with contextual metadata. 
We use a temporal split where data prior to 2022 defines the subgraph $G_{t-1}$, used for HGCR training, and edges introduced in 2022 and later define our test set at timestamp \( t \).

Each test instance is a query $q = (m_i, m_j) \in Q$, where $(m_i, m_j)$ is an experimentally verified connection that exists in $G_{t}$ (but not in $G_{t-1}$), which is also identified as a co-occurrence in the MEDLINE corpus with first appearance of this term pair in the literature at time $t$. The future reference set $FR(q)$ for a query is the collection of scientific abstracts from timestamp $t$ that mention both $m_i$ and $m_j$ (abstract-level co-occurrence).

For each query $q$, we sample candidate network paths $p \in P(q)$ from $G_{t-1}$, where each path $p_k = (m_i, \dots, m_j)$ is a shortest path between $m_i$ and $m_j$ that we limit to length 4 due to computational constraints. 
Note that multiple shortest paths in $G$ are possible.
Paths are labeled as positive or negative based on criteria described in the Methodology section. This results in a strongly imbalanced dataset which histogram is shown on Figure~\ref{fig:test_paths_len_hist}. We limit number of negative paths per query to 200 (100 paths of length 3 and 100 paths of length 4) to keep the task computationally feasible and yet this corresponds to a mean negative-to-positive ratio of approximately 20:1, though this varies across queries due to the underlying graph structure.

HGCR is evaluated as an information retrieval system over labeled paths $(p, l(p))$, while the explainability framework is assessed based on its ability to generate explanations that align with their future references $FR(q)$. For explainability evaluation, we use a subset of queries $Q_{\text{expl}} \subset Q$, selected such that each $q \in Q_{\text{expl}}$ has strong semantic connection to its future reference set, i.e., we keep only queries where both $m_i$ and $m_j$ appear together in the title of at least one future reference abstract $FR(q)$, or where a semantic predicate of the form $(m_i, \text{verb}, m_j)$ is identified via the SemRep relation extraction system \citep{rindflesch03}. This ensures that $FR(q)$ provides a meaningful reference for evaluating the generated explanations.

\subsection{Evaluation Metrics}

\subsubsection{Information Retrieval metrics}

The metrics we use to report the HGCR performance are commonly used in the information retrieval field and describe the ranking performance: the area under receiver-operating characteristic curve (AUC ROC) and average precision (AP). 

AUC ROC measures the trade-off between true positive rate and false positive rate across all thresholds, capturing the model’s ability to distinguish between positive and negative samples independently of their absolute scores. AP summarizes the precision–recall curve, and is sensitive to the ranking of positive instances making it more appropriate in highly imbalanced settings, where correctly identifying rare positives is critical.

We report both micro-averaged and macro-averaged scores. 
In our context, where the number of paths per query is highly variable and class imbalance is significant, macro-averaging better reflects consistency across queries, while micro-averaging is dominated by high-volume queries and highlights global performance.

\subsubsection{Explainability Evaluation Metrics}
\label{expl_eval_metrics} 

To evaluate the quality and factual alignment of generated explanations with novel scientific knowledge, we employ three complementary metrics. First, \emph{lexical similarity} is measured using the Jaccard Index over UMLS terms extracted from the generated explanation and the reference abstract. Second, \emph{semantic similarity} is assessed via the dot product between latent embeddings of the explanation and reference abstract, using two domain-specific models: MedCPT and SciNCL. They were selected for their strong performance in biomedical text representation learning \citep{jin2023medcpt, ostendorff2022neighborhood}. 

Finally, we report the error rate, defined as the proportion of low-ranked statements among all extracted statements in the generated explanation. This number reflects how frequently the system produces claims that are not supported by the hypothesis predictor. To compute the ranking criteria, we apply the same strategy outlined in Step 5 of the feedback loop (Section~\ref{rel_eval_w_AGATHA}) that identifies wrong statements in the generated explanations.

\section{Results}
In this section, we present the results of our experiments and report evaluation metrics for the methods described earlier. We evaluate the retriever and the explainability framework separately, each within its respective task.

\subsection{HGCR Context Ranking Evaluation}
\begin{table}[h]
    \centering
    \begin{tabular}{lllllll}
    \toprule
     &  &  & \multicolumn{2}{c}{ROC AUC} & \multicolumn{2}{c}{AP} \\
     &  &  & Micro & Macro & Micro & Macro \\
    Model & Text Encoder & Terms Encoder &  &  &  &  \\
    \midrule
    \multirow[t]{6}{*}{HGCR} & \multirow[t]{3}{*}{MedCPT} & AGATHA & \textbf{0.895} & \textbf{0.932} & \textbf{0.36} & \underline{0.536} \\
     &  & MedCPT & 0.893 & 0.927 & 0.318 & 0.519 \\
     &  & SapBERT & \underline{0.894} & 0.927 & 0.329 & 0.516 \\
    \cline{2-7}
     & \multirow[t]{3}{*}{PubMedNCL} & AGATHA & 0.891 & 0.929 & \underline{0.358} & 0.525 \\
     &  & MedCPT & 0.891 & 0.929 & 0.33 & 0.531 \\
     &  & SapBERT & 0.893 & \underline{0.929} & 0.352 & \textbf{0.536} \\
    \cline{1-7} \cline{2-7}
    MedCPT & Article Encoder & Query Encoder & 0.705 & 0.696 & 0.11 & 0.153 \\
    \cline{1-7} \cline{2-7}
    \end{tabular}
    \caption{
        Evaluation of HGCR model variants and comparison with MedCPT IR model using micro- and macro-averaged ROC AUC and Average Precision (AP). All HGCR models use the same architecture with different combinations of term and text encoders. Best values are hignlighted in \textbf{bold}, second best are \underline{underlined}. 
    }
    \label{tab:hgcr_retrieval_quality}
\end{table}

Table~\ref{tab:hgcr_retrieval_quality} summarizes the performance of HGCR across multiple encoder configurations using micro- and macro-averaged AUC and average precision metrics. The best results are consistently achieved when combining MedCPT for context encoding with AGATHA for term encoding, showing the benefit of aligning fine-tuned biomedical text representations with graph-based concept embeddings.

The HGCR model using AGATHA Term Encoder outperforms SapBERT and MedCPT variants across macro metrics, suggesting slightly better generalization to rare or diverse biomedical connections. Replacing the text encoder with PubMedNCL yields similar performance, confirming the robustness of the HGCR architecture with respect to contextual input.

We also include one baseline method based on zero-shot biomedical information retrieval model MedCPT \citep{jin2023medcpt}. It includes article and query encoders, which are utilized to compare its performance in the ranking evaluation setting. MedCPT computes query–document similarity directly without modeling paths, and we evaluate it by pooling context from positive and negative paths (ensuring that there is no overlap) and computing per-query scores. As expected, it does not perform as good as path-based models, because it lacks the ability to model multi-hop relational network structure and is tuned to work as an information retrieval system and not a predictor.

\subsection{Explainability Framework Evaluation}
We evaluate our proposed explainability approach by measuring how well the generated explanations align with corresponding future reference abstracts and how factually correct they are. To this end, we use multiple metrics described in Section~\ref{expl_eval_metrics}. The results are presented in Table~\ref{tab:comparison_with_error}.

To ensure a fair comparison across systems, we controlled the information available to each model. All queries from $Q_{\text{expl}}$ were first submitted to the VAIV system (the only system requiring interaction with a proprietary online platform\footnote{\url{https://bio.vaiv.kr/}}). After obtaining the VAIV outputs, we filtered out any query $q$ which explanation $e_{\text{VAIV}}$ included at least one abstract published in 2022 or later. This resulted in a filtered query set of 106 examples, which was then used across all systems.
For MedCPT-based retrieval systems, we restricted the document corpus to abstracts published prior to 2022, aligning with the constraints applied to HGCR.

To account for variability, we generate explanations three times for non-retrieval models (with non-zero temperature), HGCR-based retrievals (based on the top-3 scored paths per query), and VAIV (which produced different outputs for repeated queries). Scores are averaged across runs. In contrast, MedRAG and iMedRAG were executed once with default parameters (including the LLM model), as they both internally use deterministic retrieval and fixed near-zero LLM temperature.

For reference, we also report results for standalone LLMs without retrieval. However, this comparison is inherently unfair, as these models were trained on data containing the future reference abstracts as well as works that cite them or mention the same results (from 2022 onward), which is evident from the reported knowledge cutoff\footnote{\url{https://github.com/meta-llama/llama-models/blob/main/models/llama3_3/MODEL_CARD.md}}. Applying the same temporal restrictions as for retrieval systems is not feasible for pre-trained LLMs.

As shown in Table~\ref{tab:comparison_with_error}, the proposed HGCR-based approach yields explanations that better align with future references, both lexically and in latent embedding spaces. Notably, MedRAG and iMedRAG achieve relatively high MedCPT similarity despite lower Jaccard and SciNCL scores. 

The main goal of the feedback loop is to reduce hallucinations and unsupported claims, which is reflected in consistently lower error rates across models when the feedback mechanism is applied (rows marked with FL in Table~\ref{tab:comparison_with_error}). While the Llama-3.3 70B model without retrieval performs best overall in terms of error rate, the feedback loop-enabled Llama-3.1 8B closely follows. This can be explained by the ability of larger models like Llama-3.3 70B to manage its own parametric knowledge since there is likely an overlap between these models' knowledge cutoff and our test set, as was mentioned earlier. Other models equipped with the feedback loop also achieve comparable gains, demonstrating its general effectiveness in enhancing explanation reliability.
\begin{table}[!t]
  \centering
  \caption{
    Performance comparison between the proposed explainability framework and similar systems. We report average values and standard deviation across 3 runs unless specified otherwise. $(\cdot)$ represents dot product between the explanation embedding and reference abstract embedding in the corresponding latent space. Best values are highlighted in \textbf{bold}, second best are \underline{underlined}.
  }
  \Large
  \resizebox{\textwidth}{!}{%
  \begin{tabular}{lllcccc}
  \toprule
   & & & Jaccard & MedCPT & Scincl & Error Rate \\
LLM & Explainer & Retriever & Index & $(\cdot$) & $(\cdot$) & (mean ± std) \\
\midrule
\multirow[t]{3}{*}{Phi-4} 
  & Prompt & N/A & 0.070 ± 0.027 & 61.098 ± 2.884 & 488.002 ± 27.264 & 0.044 ± 0.074 \\
  & BL & HGCR & \textbf{0.085 ± 0.034} & \textbf{62.052 ± 2.991} & \textbf{497.184 ± 25.293} & 0.046 ± 0.063 \\
  & FL & HGCR & 0.080 ± 0.032 & \underline{61.982 ± 2.976} & \underline{496.042 ± 26.718} & 0.020 ± 0.044 \\
\cline{1-7}
\multirow[t]{3}{*}{Llama-3.1 8B} 
  & Prompt & N/A & 0.067 ± 0.026 & 60.573 ± 2.747 & 488.665 ± 29.073 & 0.045 ± 0.072 \\
  & BL & HGCR & 0.075 ± 0.033 & 61.514 ± 2.736 & 494.373 ± 23.093 & 0.045 ± 0.072 \\
  & FL & HGCR & 0.073 ± 0.033 & 61.169 ± 2.695 & 491.247 ± 23.344 & \underline{0.015 ± 0.043} \\
\cline{1-7}
\multirow[t]{3}{*}{Llama-3.3 70B} 
  & Prompt & N/A & 0.071 ± 0.029 & 60.711 ± 2.742 & 490.788 ± 27.637 & \textbf{0.012 ± 0.032} \\
  & BL & HGCR & \underline{0.081 ± 0.030} & 61.859 ± 2.895 & 495.750 ± 24.507 & 0.045 ± 0.073 \\
  & FL & HGCR & 0.078 ± 0.029 & 61.692 ± 2.869 & 494.702 ± 25.124 & 0.021 ± 0.064 \\
\cline{1-7}
\multirow[t]{3}{*}{ChatGPT-3.5} 
  & MedRag & MedCPT & 0.059 ± 0.029 & 61.471 ± 3.537 & 478.173 ± 34.952 & 0.079 ± 0.214 \\
  & iMedRag & MedCPT & 0.062 ± 0.024 & 61.518 ± 3.234 & 484.051 ± 31.241 & 0.060 ± 0.180 \\
  & VAIV & Custom & 0.065 ± 0.022 & 60.904 ± 3.247 & 480.932 ± 30.222 & 0.044 ± 0.061 \\
\cline{1-7}
\end{tabular}
}
  \label{tab:comparison_with_error}
\end{table}

\subsection{Case study}
The interpretability of LLM-generated explanations is critical for fostering trust in hypothesis generation systems, particularly among biomedical researchers and clinicians. Building on prior work that highlights the value of human feedback \citep{tyagin22}, we conducted an expert evaluation to assess the explanations of five pairwise concept associations proposed by domain experts in drug discovery area. After the explanations were generated, the experts reviewed them for biological plausibility and interpretability to gauge the real-world utility of our system.

For the GABRA5 $\leftrightarrow$ carbamazepine association, the explanation effectively connects two concepts by highlighting their roles in neural excitability and psychiatric conditions; Isavuconazole $\leftrightarrow$ sunitinib pair is supported by its shared interaction with CYP3A4, where isavuconazole’s inhibition of the enzyme can raise sunitinib levels and risk of toxicity, as shown by \citep{hu2024isavuconazole}; and SLC6A3 $\leftrightarrow$ bupropion, where the drug’s inhibition of the dopamine transporter explains its behavioral effects. 
Icariin $\leftrightarrow$ vascular dementia is linked through icariin’s neuroprotective effects and modulation of PI3K/Akt and MEK/ERK pathways, potentially mitigating apathy as a dopamine-related symptom via enhanced VTA–NAcc signaling, while SEZ6L2 $\leftrightarrow$ calcitriol highlights overlapping pathways in neural development and glioma biology. 
These explanations demonstrate the effectiveness of the proposed path-based RAG framework in providing targeted biomedical context, enabling LLMs to generate novel and plausible hypotheses that are consistent with expert understanding and advance scientific discovery (See Appendix~\ref{appendix:case_study} for full explanations).

\subsection{Retrieval Score and Explanation Alignment with Future Reference Context}
\label{sec:hgcr_vs_sim} 
    \begin{figure}
        \centering
        \includegraphics[
            width=0.85\linewidth,
        ]{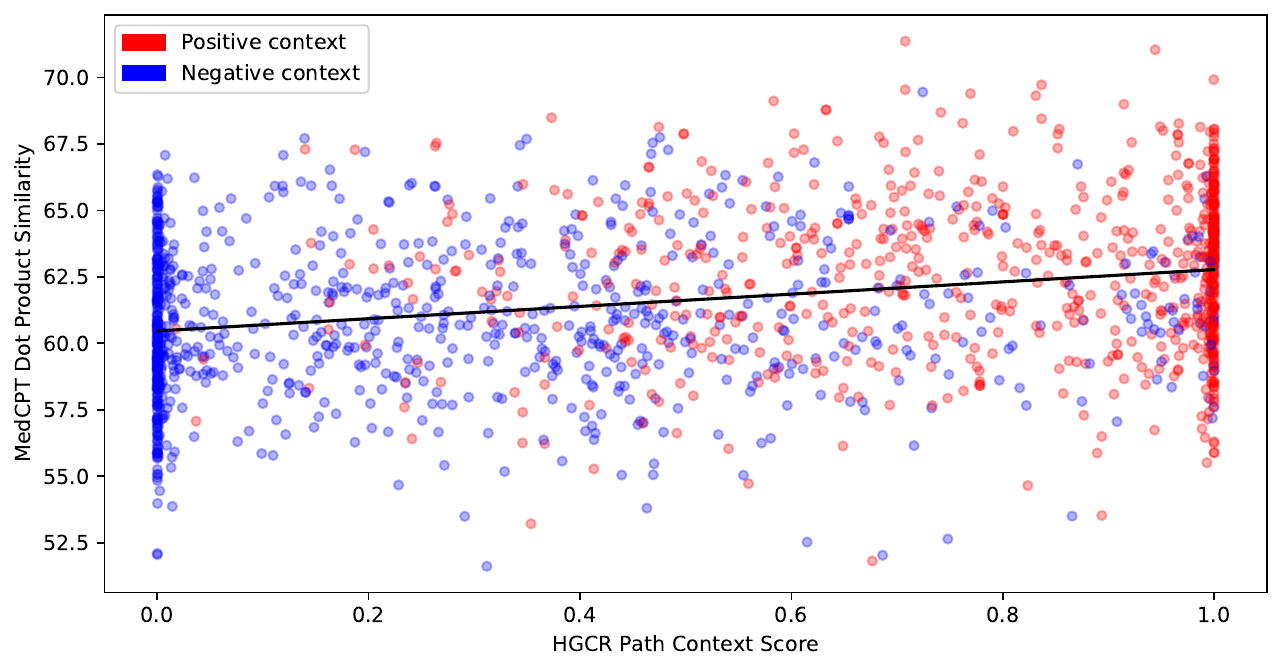}
        \caption{%
            \label{fig:hgcr_score_vs_dot_product}
            Relationship between HGCR Context Score (horizontal axis) and semantic similarity (MedCPT Dot Product, vertical axis) between final LLM-generated explanations (with feedback loop) and future reference scientific abstracts.
        }
    \end{figure}

In this section, we evaluate whether the HGCR path score reflects alignment between generated explanations and future reference abstracts. Specifically, we use a subset of queries $Q_{\text{expl}}$ and, for each query $q \in Q_{\text{expl}}$, sample an equal number of positive and negative context paths $p_k \in P^+(q) \cup P^{-}_{\text{hard}}(q)$. We fix the number of sampled paths to 3 per class.

Each path $p_k$ is assigned an HGCR context score $\hat{S}(p_k) \in [0, 1]$, computed as described in Methodology section. The explanation $e(p_k)$ corresponding to $p_k$ is generated by iteratively applying the feedback loop until convergence. We measure the semantic alignment between $e(p_k)$ and the future reference abstract $FR(q)$ using MedCPT model, computing their similarity via dot product like it is done in the previous section. 

We then analyze the relationship between $\hat{S}(p_k)$ and $sim_{\text{MedCPT}}(e(p_k), FR(q))$ across all samples, where $sim_{\text{MedCPT}}(e(p_k), FR(q))$ denotes the dot product similarity between the MedCPT embeddings of the generated explanation and the future reference abstract. Figure~\ref{fig:hgcr_score_vs_dot_product} plots this relationship for the explanations generated with Llama-3.3-70B model. Red points indicate paths sampled from positive (future-aligned) contexts, while blue points represent negative (non-aligned) contexts.
A linear regression line clearly indicates a positive correlation, which suggests that the HGCR path ranking criteria can indicate the degree of semantic alignment between generated explanations and future scientific knowledge, as captured by reference abstracts.

\subsection{Ablation Study: Context Size Influence on Explanation Quality}
To assess the effect of increasing context length on LLM performance, we analyze key evaluation metrics across different context sizes \(k \in \{1, 3, 5, 7, 9, 11\}\) for Phi-4 model. We keep the dataset the same as in previous experiment (Section~\ref{sec:hgcr_vs_sim}), but we select only positively-labeled paths to simulate best-case scenario for the language model and reduce noise. The results (Figure~\ref{fig:ablation_metrics_revision_horizontal}) indicate that increasing $k$ generally improves semantic similarity, particularly in the range $k=3$ to $k=7$. Based on these observations, the value $k=7$ was used in our main experiment in the Results section.

\begin{figure}[!t]
 \centering 
  \includegraphics[width=\textwidth]
 {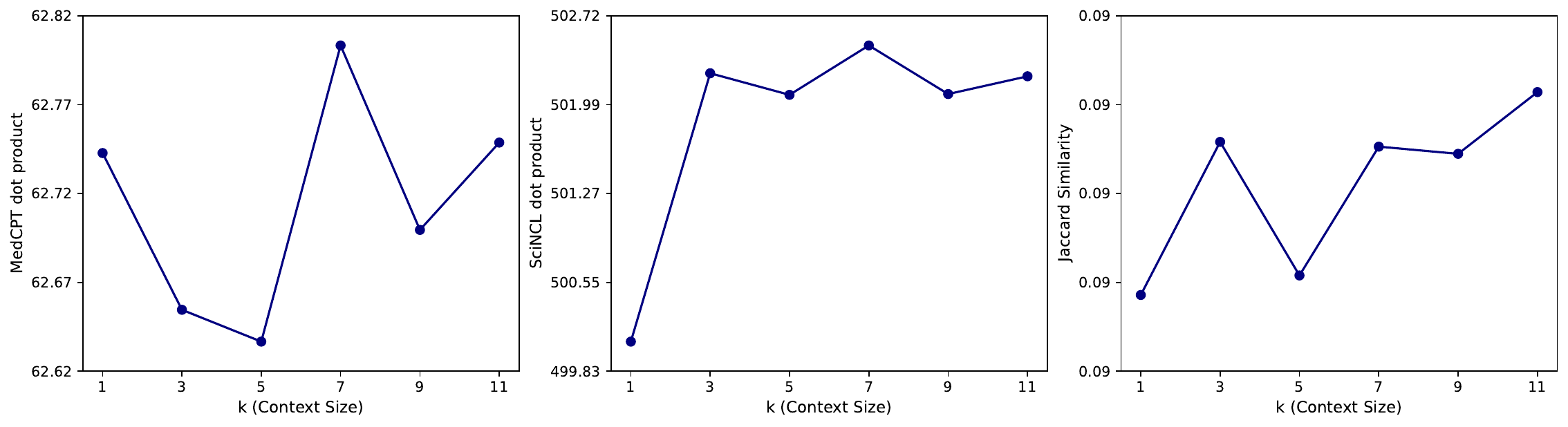}
 \caption{Different metrics across context sizes ($k$) in the ablation study. Larger context size tends to improve similarity metrics.}
 \label{fig:ablation_metrics_revision_horizontal} 
\end{figure}

\paragraph{Limitations.}
Evaluating free-text explanations is inherently difficult, especially in biomedical domains. Most prior work focuses on structured outputs \citep{xiong24biMedRag, xiong24, 10.1093/bioinformatics/btae560}, with limited attention to the quality and scientific plausibility of open-ended explanations. Moreover, identifying appropriate scientific context at scale remains challenging. This is an issue we attempt to address, but we should admit that it is still an open problem. 

We also acknowledge that large language models may use their internal knowledge beyond HGCR’s cutoffs, which became especially evident in case of Llama-3.3 70B model without any retrieval. Our primary retrieval-focused explanation evaluation strategy is supported by the results of ClashEval study \citep{wu2024clasheval} showing that LLMs tend to override their parametric knowledge with supplied context, which addresses this limitation, although not completely. Constructing the evaluation dataset with non-overlapping knowledge cutoffs between LLMs and recent scientific discoveries is not quite feasible due to the rapid models development and the nature of their training, especially using high-quality scientific data. Lastly, the feedback loop context update rule relies on local similarity-based heuristics, which may reinforce context drift or amplify biases introduced by early generation errors.

\bibliography{sample,ilya-biblio}

\newpage
\appendix

\section{Case study}
\label{appendix:case_study}
Here we provide the full version of five source–target concept pairs alongside their corresponding explanations, which were evaluated by a domain expert to assess biological plausibility and interpretability.

\subsection{GABRA5 gene and carbamazepine}
\paragraph{Explanation:}
To explore the potential indirect relationship between the GABRA5 gene and carbamazepine, we need to consider the information provided in the abstracts and connect the dots through genetic, pharmacological, and neurobiological pathways.
GABRA5 Gene and Suicidal Behavior: The first abstract discusses genetic variants associated with suicidal behavior, including the GABRA5 gene. GABRA5 is part of the GABA A receptor complex, which is involved in inhibitory neurotransmission in the brain. Variants in this gene have been linked to psychiatric conditions and behaviors, including those related to suicide risk. Carbamazepine and Seizure Control: The third abstract mentions the use of carbamazepine in treating seizures in a patient with Dyke-Davidoff-Masson syndrome (DDMS). Carbamazepine is an anticonvulsant and mood-stabilizing drug that works by stabilizing hyperexcited nerve membranes, inhibiting repetitive neuronal firing, and reducing synaptic propagation of excitatory impulses. It primarily acts on sodium channels but also has effects on GABAergic neurotransmission. Indirect Link via Neurotransmission: The connection between GABRA5 and carbamazepine can be hypothesized through their roles in neurotransmission. GABRA5, as part of the GABA A receptor, influences inhibitory signaling in the brain. Carbamazepine, while primarily affecting sodium channels, may also modulate GABAergic activity indirectly. This modulation could influence the overall excitatory-inhibitory balance in the brain, which is crucial in conditions like epilepsy and potentially in mood disorders linked to suicidal behavior.
Potential Intermediate Mechanisms: The interaction between GABRA5 variants and carbamazepine's effects might involve changes in GABAergic signaling pathways. If GABRA5 variants alter GABA receptor function, this could affect how carbamazepine modulates neuronal excitability and seizure control. Additionally, since GABAergic dysfunction is implicated in various psychiatric disorders, carbamazepine's influence on this system might also impact mood and behavior, indirectly linking it to the genetic risk factors for suicidal behavior.
Gene-Environment Interactions: The first abstract also mentions gene-environment interactions, such as the interaction between NTRK2 genetic variation and childhood abuse. While not directly related to GABRA5 or carbamazepine, this highlights the complexity of genetic influences on behavior and how environmental factors can modulate genetic risk, potentially affecting how individuals respond to medications like carbamazepine.
In summary, the indirect relationship between the GABRA5 gene and carbamazepine may exist through their shared involvement in modulating neurotransmission, particularly GABAergic signaling. Variants in GABRA5 could influence the efficacy or side effects of carbamazepine by altering the inhibitory-excitatory balance in the brain, which is crucial for both seizure control and mood regulation. Further research would be needed to elucidate these pathways and their clinical implications.
\paragraph{Expert assessment:}
Explanation correrctly identifies the role of GABRA5 in GABAergic neurotransmission and links it to carbamazepine's effects on neuronal excitability. Additionally, it considers gene environment interactions and psychiatric implications, providing a more comprehensive view of their potential connection. The GABRA5 gene encodes the $\alpha$5 subunit of the GABA A receptor, a key component of inhibitory neurotransmission in the brain. It plays a role in neurodevelopment, synaptic plasticity, and psychiatric disorders. Carbamazepine is an anticonvulsant and mood stabilizer that primarily acts by blocking voltage gated sodium channels, reducing neuronal excitability. It is commonly used to treat epilepsy, bipolar disorder, and trigeminal neuralgia. The indirect relationship between GABRA5 and carbamazepine likely involves their shared involvement in neural excitability and plasticity. GABRA5 regulates inhibitory neurotransmission, while carbamazepine modulates excitatory activity by stabilizing sodium channels. Through the balance of excitatory inhibitory signaling, alterations in GABRA5 expression or function could influence carbamazepine’s effectiveness or mechanisms of action.

\subsection{SLC6A3 gene and bupropion}
\paragraph{Explanation:}
To explore the potential indirect relationship between the SLC6A3 gene and bupropion, we need to consider the roles and interactions of these elements based on the provided abstracts.
SLC6A3 Gene: The SLC6A3 gene encodes the dopamine transporter (DAT), which is crucial for regulating dopamine levels in the brain by facilitating the reuptake of dopamine from the synaptic cleft back into presynaptic neurons. This gene is part of the dopaminergic system, which is implicated in cognitive processes and intelligence, as mentioned in the first abstract.
Bupropion: Bupropion is an antidepressant and smoking cessation aid that primarily acts as a norepinephrine-dopamine reuptake inhibitor (NDRI). By inhibiting the reuptake of dopamine and norepinephrine, bupropion increases the levels of these neurotransmitters in the synaptic cleft, enhancing dopaminergic and adrenergic signaling.
Indirect Relationship:
Dopaminergic System: The SLC6A3 gene, through its product (DAT), plays a role in modulating dopamine levels. Bupropion, by inhibiting DAT, indirectly affects the function of the SLC6A3 gene product, leading to increased synaptic dopamine levels.
Cognitive and Behavioral Effects: The dopaminergic system, influenced by SLC6A3, is involved in cognitive processes and mood regulation. Bupropion's action on dopamine reuptake can alter these processes, potentially impacting cognitive functions and mood, which are areas where SLC6A3 is also relevant.
Side Effects and Psoriasis: The third abstract highlights a case where bupropion exacerbated erythrodermic psoriasis, a severe skin condition. While this is not directly related to SLC6A3, it underscores the complex interactions and side effects that can arise from altering neurotransmitter systems, which are influenced by genes like SLC6A3. In summary, the indirect relationship between the SLC6A3 gene and bupropion can be understood through their shared involvement in the dopaminergic system. Bupropion affects dopamine levels by inhibiting the dopamine transporter encoded by SLC6A3, thereby influencing cognitive and mood-related processes. This interaction highlights the broader implications of pharmacological interventions on genetic and neurobiological pathways.

\paragraph{Expert assessment:}
Explanation offers a particularly direct and relevant connection between the SLC6A3 gene and bupropion. It focuses on their shared role in the dopaminergic system, clearly explaining how the gene encodes the dopamine transporter (DAT) and how bupropion inhibits DAT, leading to increased synaptic dopamine levels. This direct mechanism is central to understanding their interaction. This connection is further supported by its logical extension to cognitive and behavioral effects, which are well-established areas of influence for both dopamine regulation and bupropion. It also touches on the potential for side effects, showing a more comprehensive view.

\subsection{Isavuconazole and Sunitinib}
\paragraph{Explanation:}
An indirect relationship between isavuconazole and sunitinib can be proposed based on the provided abstracts, focusing on the role of CYP3A inhibitors and their impact on the pharmacokinetics of both drugs. Here's a coherent narrative illustrating this possible indirect linkage:
Isavuconazole and CYP3A Interaction: Isavuconazole is known to be a moderate inhibitor of the CYP3A enzyme. Although the abstracts do not directly discuss isavuconazole's interaction with sunitinib, understanding its role as a CYP3A inhibitor is crucial.Sunitinib Metabolism and CYP3A: Sunitinib is primarily metabolized by the CYP3A enzyme. The study on the effects of CYP3A inhibitors (ketoconazole, voriconazole, and itraconazole) on the pharmacokinetics of sunitinib shows that strong CYP3A inhibitors like ketoconazole and voriconazole can significantly increase the exposure of sunitinib by inhibiting its metabolism. This suggests that any drug that inhibits CYP3A could potentially affect sunitinib's pharmacokinetics.Indirect Relationship: Given that isavuconazole is a CYP3A inhibitor, albeit moderate, it could potentially influence the pharmacokinetics of sunitinib if co-administered. The mechanism would involve isavuconazole inhibiting the CYP3A enzyme, which in turn could reduce the metabolism of sunitinib, leading to increased exposure of sunitinib in the body. This increased exposure could enhance the efficacy of sunitinib but also potentially increase the risk of toxicity.
Intermediate Entities or Processes: The key intermediate entity in this relationship is the CYP3A enzyme. The process involves the inhibition of this enzyme by isavuconazole, which then affects the metabolism of sunitinib. Another intermediate process could be the alteration in the pharmacokinetic profile of sunitinib, including its absorption, distribution, metabolism, and excretion (ADME), primarily due to the inhibition of its metabolism.
Clinical Implications: Understanding this indirect relationship is crucial for clinical practice. If isavuconazole and sunitinib are to be used concomitantly, healthcare providers should be aware of the potential for increased sunitinib exposure and monitor patients closely for signs of toxicity or enhanced efficacy. This might necessitate dose adjustments of sunitinib or careful monitoring of its levels to maintain therapeutic efficacy while minimizing adverse effects. In summary, the indirect relationship between isavuconazole and sunitinib is mediated through their interaction with the CYP3A enzyme. Isavuconazole, as a CYP3A inhibitor, can potentially increase the exposure of sunitinib by reducing its metabolism, thereby indirectly affecting its pharmacokinetics and potentially its efficacy and safety profile.

\paragraph{Expert assessment:}
The link between isavuconazole and sunitinib has the potential to be true because both drugs interact with the same enzyme, CYP3A4. Isavuconazole is a moderate inhibitor of CYP3A4, which slows down the enzyme’s activity. Sunitinib, on the other hand, relies almost entirely on CYP3A4 to break down and leave the body. It is known that strong CYP3A4 inhibitors like ketoconazole can double sunitinib levels in the blood by blocking its metabolism and increasing its levels, which was shown recently by \citep{hu2024isavuconazole}. This could mean stronger effects from sunitinib and also a higher risk of side effects like high blood pressure, fatigue, or liver dysfunction. It is important to monitor patients closely and adjust the dose of sunitinib if the two are co-administered.

\subsection{Icariin and Dementia, Vascular}
\paragraph{Explanation:}
Based on the provided abstracts, an indirect relationship between icariin and vascular dementia (VaD) can be proposed through the following mechanisms and intermediate entities: Icariin's neuroprotective effects: Icariin has been shown to have neuroprotective properties, particularly in the context of Parkinson's disease (PD). It can protect dopaminergic neurons and modulate signaling pathways such as PI3K/Akt and MEK/ERK, which are involved in cell survival and neuroprotection. Dopamine pathway and apathy: The studies on MPTP-lesioned primates suggest that dopaminergic dysfunction, particularly in the ventral tegmental area (VTA)-nucleus accumbens (NAcc) pathway, contributes to the manifestation of apathetic behaviors. Apathy is a common symptom in both PD and VaD. Apathy and social withdrawal in VaD: The study on social withdrawal in patients with VaD found that apathy, impaired social judgment, and disturbance of basic daily activities are associated with social withdrawal. Apathy, in particular, was a significant predictor of social withdrawal. Link between icariin and VaD: While there is no direct evidence of icariin's effects on VaD, its neuroprotective properties and potential to modulate dopamine pathways could indirectly influence the development or progression of VaD. Icariin might help mitigate apathy and social withdrawal in VaD patients by protecting dopaminergic neurons and promoting dopamine signaling. The proposed indirect relationship between icariin and VaD can be summarized as follows:
Icariin → Neuroprotection of dopaminergic neurons → Modulation of dopamine pathways (e.g., VTA-NAcc) → Reduction of apathy → Decreased social withdrawal → Potential improvement in VaD symptoms
This hypothetical relationship highlights the potential for icariin to influence VaD symptoms indirectly, through its effects on dopamine pathways and apathy. However, further research is necessary to explore this relationship and determine the therapeutic potential of icariin in VaD.

\paragraph{Expert assessment:}
This proposed connection is logical as a hypothesis because it links established properties of icariin with a key symptom of vascular dementia through a shared biological pathway. It is known that icariin demonstrates neuroprotective effects, particularly on dopaminergic neurons, and can modulate critical cell survival pathways like PI3K/Akt and MEK/ERK. The apathy, which is one of the debilitating symptoms in vascular dementia, is understood to involve dysfunction in dopamine signaling, specifically pathways like the VTA-NAcc connection. Therefore, it is biologically plausible that icariin's ability to protect dopamine neurons and enhance dopamine-related signaling could indirectly improve vascular dementia symptoms by mitigating apathy and its downstream consequences, like social withdrawal. 

\subsection{SEZ6L2 gene and calcitriol}
\paragraph{Explanation:}
To explore the potential indirect relationship between the SEZ6L2 gene and calcitriol, we need to consider the roles and interactions of these entities within the context of neural development and glioma biology, as described in the provided abstracts.
Key Findings and Mechanisms:
SEZ6L2 Gene in Neural Development:
The SEZ6L2 gene is identified as differentially expressed during mouse neural tube development, with alternative splicing observed. This suggests its involvement in the complex regulation of neural development processes.
SEZ6L2 is among the genes dysregulated in a retinoic acid-induced neural tube defects (NTDs) model, indicating its potential role in neural development and associated pathologies.
Calcitriol in Glioma Biology:
Calcitriol, the active form of vitamin D3, promotes differentiation of glioma stem-like cells (GSCs) and increases their susceptibility to temozolomide, a common chemotherapeutic agent for glioblastoma.
The study suggests that calcitriol can reduce the stemness of GSCs, potentially impacting tumor growth and treatment resistance.
Indirect Relationship:
To hypothesize an indirect relationship between SEZ6L2 and calcitriol, we can consider the following narrative:
Neural Development and Glioma Initiation:
SEZ6L2's role in neural development suggests it may influence the cellular environment or pathways that are also relevant in glioma initiation or progression. Neural progenitors and stem cells, which are implicated in both normal neural development and glioma formation, could be a connecting point.
Calcitriol's Impact on Glioma Stem Cells:
Calcitriol affects glioma stem-like cells by promoting differentiation and reducing stemness, which could alter the cellular landscape in a way that impacts genes involved in neural development, such as SEZ6L2.
Potential Intermediate Pathways:
Both SEZ6L2 and calcitriol may interact with common signaling pathways or cellular processes, such as those involving transcription factors or signaling molecules that regulate cell differentiation and proliferation.
For instance, pathways involving retinoic acid (mentioned in the context of SEZ6L2 dysregulation) and vitamin D signaling (related to calcitriol) could intersect, influencing gene expression profiles and cellular behaviors in neural and glioma contexts.
Conclusion:
While the abstracts do not provide direct evidence of an interaction between SEZ6L2 and calcitriol, the indirect relationship could be mediated through shared pathways in neural development and glioma biology. Further research would be needed to elucidate specific molecular mechanisms and validate this hypothesis.
\paragraph{Expert assessment:}
A notable strength of the explanation lies in a tight focus on the neural context, directly addressing the roles of the SEZ6L2 gene in neural development and calcitriol in glioma biology. By meticulously outlining the potential connections through neural progenitor cells and shared signaling pathways, it provides a logical and coherent description. The explanation doesn't just state a possibility; it researches into potential mechanisms, such as the interplay between retinoic acid and vitamin D signaling, demonstrating a deeper understanding of the subject matter. Furthermore, it strikes a balance between acknowledging the absence of direct evidence and presenting a plausible hypothesis, making it a well-reasoned and insightful response.

\section{Incorporated Technologies} 
The proposed pipeline relies on multiple key technologies and biomedical databases to conduct experiments effectively.

\subsection{UMLS (Unified Medical Language System)}  
UMLS \citep{bodenreider04}, developed by the National Library of Medicine (NLM), integrates and standardizes biomedical terminologies, taxonomies, and coding systems. It provides concept unique identifiers (CUIs) that unify terms from diverse vocabularies, facilitating interoperability across biomedical resources.

\subsection{MetaMap and SemRep}  
MetaMap \citep{aronson01} is an NLM-developed software that identifies biomedical concepts in text using Named Entity Recognition (NER) and maps them to UMLS concepts, enabling standardized representation. SemRep \citep{rindflesch03}, also developed by NLM, leverages MetaMap’s capabilities to extract structured semantic predicates from biomedical literature, representing relationships as subject-verb-object predicates and capturing explicit biomedical knowledge through these entities.


\subsection{AGATHA}
AGATHA \citep{agathasybrandt20} (Automatic Graph-mining And Transformer-based Hypothesis generation Approach) is a versatile hypothesis generation (HG) system that integrates a multi-layered semantic graph with transformer-based deep learning techniques. It constructs a large-scale semantic network from biomedical literature, applies advanced NLP techniques, and utilizes transformer-based architecture to predict and rank plausible connections efficiently. Unlike traditional link prediction systems, AGATHA offers a comprehensive framework where predicting links is just one part of a broader pipeline, making it suitable for diverse hypothesis generation tasks across biomedical literature.

\subsection{BERT-derivative Models}  
The pipeline employs BERT-based models such as SciNCL/PubMedNCL \citep{ostendorff2022neighborhoodcontrastivelearningscientific} and MedCPT~\citep{Jin_2023}, which are optimized for scientific and biomedical text processing, respectively.

\subsection{VLLM}  
VLLM \citep{kwon2023efficient} is an optimized inference and serving framework for large language models (LLMs), enabling efficient model execution with reduced memory overhead. It supports fast decoding and inference for transformer-based models, making it particularly useful for large-scale biomedical text processing.

\section{Technical details for Large Language Models}

For the feedback loop experiment, various Large Language Models (LLMs) were utilized to generate hypotheses. To ensure fairness and consistency across experiments, we uniformly used Phi-4 \citep{abdin24}, Llama-3.1 8B \citep{grattafiori24}, and Llama-3.3 70B for all LLM-related components, including both the baseline and feedback loop experiments. For consistency and reproducibility in the feedback loop experiment, the temperature parameter was set to \( 1 \times 10^{-19} \), and the top-p value was set to \( 1 \times 10^{-9} \). These extreme values were selected to minimize randomness and ensure deterministic output across multiple iterations, helping maintain a high degree of consistency in the LLM responses. This was particularly important for iteratively refining hypotheses based on feedback.

Models Llama-3.1 8B with 8 billion parameters, Llama-3.3 70B with 70 billion parameters, and Phi-4 with 14 billion parameters were downloaded and deployed locally using VLLM~\citep{kwon2023efficient}, enabling efficient parallel processing and inference for hypothesis generation.

In our entire experiment, the prompts were designed to use a zero-shot approach, where the LLMs generated hypotheses without any task-specific fine-tuning. Furthermore, the max\_completion\_tokens parameter was set to 1000 to limit the response length during inference using VLLM, and the following parameters were applied across all models: max-model-len was set to 16384, temperature was set to \( 1 \times 10^{-19} \) and top-p was set to \( 1 \times 10^{-9} \) to ensure coherence and reduce variability. For the prompt experiment, where no retriever was used, we set the temperature to 0.5 and top-p to 0.6.


\section{Explainability Correctness Evaluation}

In addition to our main results, we attempt to evaluate factual correctness using an LLM-as-a-judge approach~\citep{chiang2023can} implemented with LlamaIndex’s \texttt{Correctness Evaluator}~\citep{Liu_LlamaIndex_2022}, which scores contextual relevance and accuracy on a 1–5 scale. This score is intended to capture both terminological overlap and deeper semantic and factual alignment. To reduce bias, we incorporate 3 strong local LLMs serving as evaluation experts: Llama-3.3 70B, Gemma 3 27B~\citep{team2025gemma} and Qwen 3 32B~\citep{qwen3}. We calculate correctness for each explanation independently and then take the average of 3 to get the final score.

The result of this experiment is shown in Table~\ref{tab:llm_correctness_scores}. Scores indicate that the best factual correctness is achieved with models that do not use any retrieval. This can be explained by the limitation we mentioned earlier: applying the appropriate temporal knowledge cutoff restrictions as we did for tested retrieval systems is not feasible for pre-trained LLMs and their exposure to our reference texts from the test set potentially gave them an unfair advantage. 
Also, we hypothesize that since LLMs rely on latent knowledge beyond the provided context, the observed difference in the correctness scores could be  a result of the epistemic uncertainty impact.

\begin{table}[h]
    \centering
    \begin{tabular}{lccc}
    \toprule
    LLM & Prompt & FL & BL \\
    \midrule
    Phi-4 & 3.874 ± 0.403 & 3.808 ± 0.566 & 3.795 ± 0.561 \\
    Llama 3.1\_8B & 3.541 ± 0.563 & 3.291 ± 0.672 & 3.360 ± 0.663 \\
    Llama 3.3\_70B & 3.996 ± 0.413 & 3.666 ± 0.624 & 3.650 ± 0.633 \\
    \bottomrule
    \end{tabular}
    \caption{
        Mean ± standard deviation of correctness scores across different models and experiments.
    }
    \label{tab:llm_correctness_scores}
\end{table}

\section{Prompt Complexity and Operational cost of Pipeline}
\begin{table}[h]
  \centering
  \caption{
    Mean response time and token statistics (Mean ± std) across different models and experiments. 
  }
  \resizebox{\textwidth}{!}{%
\begin{tabular}{llcccc}
\toprule
Model & Experiment & Mean Response Time (sec) & Total Input Tokens & Total Output Tokens & Iterations \\
\midrule
\multirow{3}{*}{Phi-4} 
  & Prompt & 8.18 ± 0.87 & 74.17 ± 7.09 & 567.34 ± 57.34 & 1 \\
  & FL  & 67.89 ± 59.66 & 11598.92 ± 8423.15 & 1215.65 ± 894.46 & 1.93 ± 1.39 \\
  & BL  & 10.72 ± 2.55 & 6273.74 ± 1623.09 & 599.45 ± 76.08 & 1 \\
  
\midrule
\multirow{3}{*}{Llama-3.1 8B}
  & Prompt & 6.47 ± 1.32 & 74.15 ± 7.06 & 672.68 ± 138.58 & 1 \\
  & FL  & 76.04 ± 74.22 & 11944.56 ± 8325.30 & 1455.44 ± 1056.21 & 2.03 ± 1.39 \\
  & BL  & 8.24 ± 2.95 & 6268.51 ± 1620.32 & 684.48 ± 136.35 & 1 \\
  
\midrule
\multirow{3}{*}{Llama-3.3 70B}
  & Prompt & 37.02 ± 4.65 & 74.15 ± 7.06 & 687.21 ± 87.33 & 1 \\
  & FL  & 117.90 ± 91.48 & 11719.58 ± 8169.37 & 1198.43 ± 838.37 & 1.97 ± 1.37 \\
  & BL  & 35.75 ± 4.69 & 6268.51 ± 1620.32 & 593.94 ± 81.20 & 1 \\
  
\bottomrule
\end{tabular}
  }
  \label{tab:response_tokens_table}
\end{table}

Prompt complexity and operational cost of the proposed pipeline are presented in Table~\ref{tab:response_tokens_table}.
 We report mean values for response times as well as input and output tokens per model and experiment type.
\section{Convergence Statistics of the Feedback Loop Experiment}
The convergence report, which quantifies how often the feedback loop stops early and how often it hits the iteration limit, is shown in Table~\ref{tab:iterations_convergence}. In this table, the total number of runs is 318 for every language model, which represents 106 queries that were selected for the results section (Table~\ref{tab:comparison_with_error}) and run with 3 different context paths.

\begin{table}[h]
  \centering
  \caption{
    Iteration counts required for convergence across models. The last row indicates cases that required more than 5 iterations and did not converge.
  }
  \resizebox{0.75\textwidth}{!}{%
\begin{tabular}{lccc}
\toprule
Iterations completed and converged & Phi-4 & Llama-3.1 8B & Llama-3.3 70B \\
\midrule
1 & 185 & 171 & 177 \\
2 & 61  & 60  & 61  \\
3 & 21  & 34  & 27  \\
4 & 11  & 14  & 18  \\
5 & 5   & 13  & 11  \\
\multicolumn{1}{l}{\textgreater 5 (did not converge, more iterations needed)} & 35  & 26  & 24  \\
\bottomrule
\end{tabular}
  }
  \label{tab:iterations_convergence}
\end{table}

\end{document}